\documentclass[aps,twocolumn,prl,floatfix,superscriptaddress,showpacs,longbilbiography]{revtex4-1}
\usepackage[utf8]{inputenc} 

\usepackage{graphicx}
\usepackage{dcolumn}
\usepackage{bm}

\usepackage{braket}
\usepackage{amsfonts}
\usepackage{amsmath}
\usepackage{amssymb}

\usepackage{color,soul}
\usepackage{wasysym}
\usepackage{mathrsfs}
\usepackage{times}

\usepackage[dvipsnames,svgnames,table]{xcolor}

\usepackage{hyperref}
\usepackage{fancyhdr}
\usepackage{float}

\usepackage[english]{babel}

\usepackage[caption=false]{subfig}
\usepackage{braket}

\hypersetup{
pdfstartview={FitH},
colorlinks=true,    
linkcolor=NavyBlue, 
citecolor=Maroon,   
filecolor=NavyBlue, 
urlcolor=NavyBlue   
}

\newcommand{\bea}{\begin{eqnarray}}
\newcommand{\eea}{\end{eqnarray}}

\newcommand{\overbar}[1]{\mkern 1.5mu\overline{\mkern-1.5mu#1\mkern-1.5mu}\mkern 1.5mu}

\begin{document}

\title{Spin-polarized tunable photocurrents}

\author{Matías Berdakin}
\thanks{These two authors contributed equally.}
\affiliation{INFIQC (CONICET-UNC), Ciudad Universitaria, Pabellón Argentina, 5000 Córdoba, Argentina.}
\affiliation{Departamento de Química Teórica y Computacional, Fac. de Ciencias Químicas, Universidad Nacional de Córdoba, Ciudad Universitaria, Pabellón Argentina, X5000HUA Córdoba, Argentina.}
\author{Esteban A. Rodríguez-Mena}
\thanks{These two authors contributed equally.}
\author{L. E. F. Foa Torres}
\affiliation{Departamento de F\'{\i}sica, Facultad de Ciencias F\'{\i}sicas y Matem\'aticas, Universidad de Chile, Santiago, Chile}

\begin{abstract}
Harnessing the unique features of topological materials for the development of a new generation of topological based devices is a challenge of paramount importance. Using Floquet scattering theory combined with atomistic models we study the interplay between laser illumination, spin and topology in a two-dimensional material with spin-orbit coupling. Starting from a topological phase, we show how laser illumination can selectively disrupt the topological edge states depending on their spin. This is manifested by the generation of pure spin photocurrents and spin-polarized charge photocurrents under linearly and circularly polarized laser-illumination, respectively. Our results open a path for the generation and control of spin-polarized photocurrents. 
\end{abstract}

\date{\today}
\maketitle

\textit{Introduction.--} The early theoretical proposals~\cite{kane_quantum_2005,kane_z_2_2005,bernevig_quantum_2006,fu_topological_2007} and the subsequent experimental realization of topological insulators~\cite{konig_quantum_2007,hsieh_topological_2008} have lined up the relentless scientific efforts of an ever growing community in Physics, Materials Science and Chemistry~\cite{ren_topological_2016,kong_opportunities_2011}. Besides interesting features such as spin-momentum locking~\cite{ortmann_topological_2015}, topologically protected states are attractive because, unlike the usual electronic states in solids, they enjoy an intrinsic robustness to perturbations and disorder. But this lack of fragility opens up new challenges for their manipulation. Typical schemes such as surface functionalization~\cite{kong_opportunities_2011} are quite ineffective when applied to topological insulators. The difficulty to cleanly turn-off conduction of charge and spin has motivated proposals for a topological field effect transistor~\cite{qian_quantum_2014,liu_spin-filtered_2014,pan_electric_2015,vandenberghe_imperfect_2017} where the electric field drives a topological transition to a trivial insulating phase, a concept that has been experimentally realized recently~\cite{collins_electric-field-tuned_2018}.

Another stream of research has been looking to exploit light-matter interaction in materials to control their electrical properties. This includes generating effects such as dichroism~\cite{xiao_valley-contrasting_2007,cao_valley-selective_2012}, a situation where electrons at different valleys absorb left and right-handed photons differently, which is of much relevance in the context of two-dimensional materials~\cite{sie_valley-selective_2015,zeng_valley_2012,zhang_generation_2014}. A different approach is aimed at using intense laser illumination to actually change the properties of the material~\cite{noauthor_quantum_2020}. Indeed, strong illumination has been demonstrated to produce hybrid electron-photon states~\cite{wang_observation_2013,mahmood_selective_2016} (Floquet-Bloch states) which may present new topological properties~\cite{oka_photovoltaic_2009,lindner_floquet_2011,rudner_band_2020} (see also Ref.~\cite{giustino_2020_2020}) and even exhibit a light-induced Hall effect~\cite{mciver_light-induced_2020}.

Here, we study laser-illumination on graphene with spin-orbit coupling and a sublattice-symmetry breaking potential. The parameters are fixed so that, in absence of radiation, the system is in a topologically insulating phase with counter-propagating spin-polarized states protected by time-reversal symmetry. Previous related studies have focused on the rich phase diagram of Floquet topological phases under strong high-frequency radiation ($\hbar\Omega$ larger than the bandwidth)~\cite{ezawa_photoinduced_2013} and also considering resonant processes~\cite{lopez_photoinduced_2015}.
In contrast to those studies, here we focus on using light to gently disrupt the native topological states. In this regime one might expect an interesting interplay between symmetry breaking (inversion symmetry, or time-reversal symmetry which can be broken or preserved by circular or linearly polarized light), spin-orbit coupling which also intertwined the valley and spin degrees of freedom, and photon-induced processes. 
Specifically, we show that laser-illumination leads to: (i) pure spin currents under linearly polarized light, and (ii) spin polarized charge currents under circular polarization. In both cases the spin (i) and charge (ii) currents flow even at zero-bias voltage. We rationalize these pumping currents in terms of a selective hybridization of electron-photon states which is enriched by valley and spin-selective selection rules under circular polarization.

\begin{figure}[hbt!]
\centering
\includegraphics[width=1.0\linewidth]{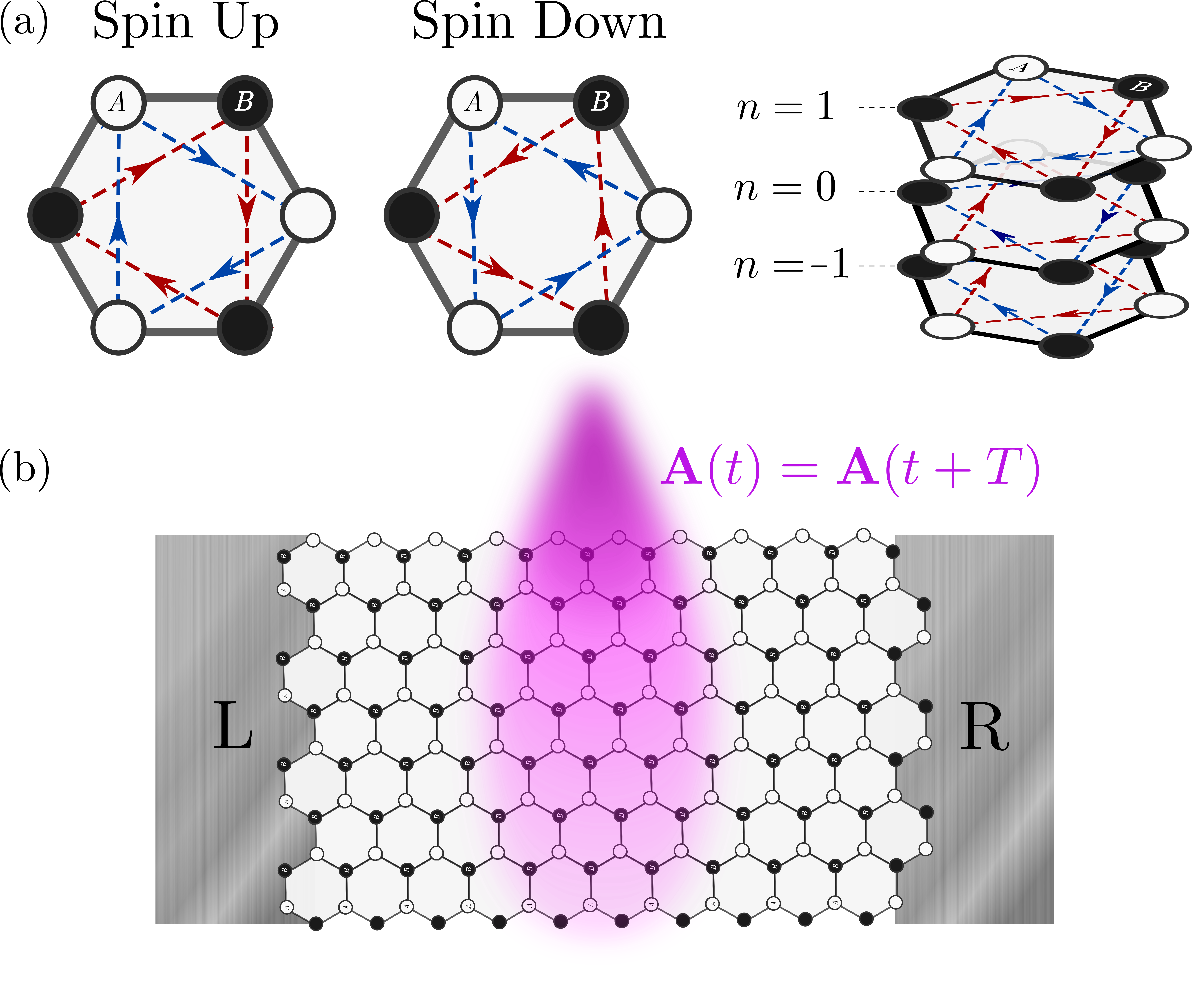}
\caption{The irradiated Kane-Mele model. The system consists of two decoupled copies, each one representing spin up and spin down, and therefore they are time-reversal partners (a, left). Under the light spot, the system develops the replica scheme unfolding itself into several copies which represent photon dressed processes (a, right). In (b) a schematic representation of the device we will consider in the transport setup. Under particular conditions the transport of one spin might be suppressed while the remaining is \emph{perfectly} unaffected.}
\label{fig:schematic}
\end{figure}

In the following we present the lattice model we use to test our ideas, the basics of Floquet theory and the generalized Landauer-Büttiker formalism. Later on we present our results for the spectral and transport properties, rationalizing them in terms of a few main ingredients. Finally, we discuss possible realizations and potential application of these ideas as well as drawbacks that may arise on the realization of these concepts.

\textit{Hamiltonian model for Floquet-Kane-Mele system.--} Let us now consider the Hamiltonian for graphene with a staggering potential and intrinsic spin-orbit (ISO) interaction~\cite{kane_quantum_2005}:
\begin{eqnarray}
\label{HamiltonianSO}
&&{\cal H}_0=\sum_{i, s_z}E_{i}^{{}}\,c_{i, s_z}^{\dagger}c_{i, s_z}^{{}}-\gamma_{0}\sum_{\left\langle i,j\right\rangle, s_z} c_{i, s_z} ^{\dagger}c_{j, s_z}^{{}}-\nonumber \\
&&-{\rm i}\gamma_{SO}\sum_{\left\langle\left\langle i,j\right\rangle\right\rangle, s_z }\nu_{i,j} s_z c_{i, s_z} ^{\dagger}c_{j, s_z}^{{}},
\end{eqnarray}
where $c_{i, s_z}^{\dagger}$ and $c_{i, s_z}^{{}}$ are the creation and annihilation operators for electrons at the $\pi$-orbital on site
$i$ with spin up $s_z=1$ or spin down $s_z=-1$. $\gamma_{0}$ is the nearest-neighbors matrix element and $\gamma_{SO}$ is the intrinsic spin-orbit coupling. We set $\gamma_0=1$ as our energy scale. $\nu_{i,j}$ is $+1$ ($-1$) if the path from $j$ to $i$ is clockwise (anticlockwise), as shown in Fig. \ref{fig:schematic}a (right for spin up, left for down). The on-site energies $E_{i}$ are chosen equal to $\Delta$ ($-\Delta$) for the sites on the A (B) sublattice. The single and double brackets denote that the summation is over first or next nearest-neighbors. Although the spin-orbit coupling in bare graphene is too small, the same physics can be realized in other two-dimensional materials such as silicene and germanene~\cite{ezawa_monolayer_2015} where this coupling is stronger.

The effect of laser illumination is captured through the Peierls' substitution~\cite{calvo_tuning_2011,calvo_non-perturbative_2013}, a time dependent phase in the nearest-neighbors and next-nearest-neighbors matrix elements:

\begin{equation}
\label{gammat}
{\gamma_{i,j}^{}(t)} = \gamma_{i,j}^{(0)} \exp \left[ {\rm i} \, \frac{2\pi}{\Phi_{0}} \int_{\mathbf{r}_j}^{\mathbf{r}_i} \mathbf{A}(t) \cdot \ d\mathbf{r} \right].
\end{equation}

\noindent where $\gamma_{i,j}^{(0)}$ are the unperturbed matrix elements as given in Eq.~\ref{HamiltonianSO}, ${\Phi_{0}}$ is the magnetic flux quantum and $\mathbf{A}(t)$ is the vector potential. For a monochromatic plane wave in the $z$-direction (perpendicular to the graphene sheet) we consider $\mathbf{A}(t)=A_0\cos(\Omega t)\hat{x}+A_0\sin(\Omega t + \phi)\hat{y}$, where $\Omega$ is the radiation frequency, $A_{0}$ determines the driving amplitude and $\phi=0,\pm\pi/2$ controls the polarization linear or left/right hand polarized, respectively. Right hand polarization is considered whenever we mention circular polarization. The laser strength can be characterized by the dimensionless parameter $z = A_{0} a 2 \pi / \Phi_0$.

Similar systems were considered before with a few differences: Ref.~\cite{huaman_floquet_2019} studied laser-illuminated transition metal dichalcogenide without spin-orbit coupling considered here, and Refs.~\cite{tahir_floquet_2016,lopez_photoinduced_2015} studied germanene and silicene in the high-frequency regime while here we focus on frequencies smaller than the bandwidth. Other studies using Floquet theory focused on the topological states induced by light~\cite{bajpai_how_2020,farrell_photon-inhibited_2015,sato_light-induced_2019}, rather than the modification of native topological states considered here.

\textit{Floquet theory for the spectral and transport properties.--} 

Floquet theory allows for a non-perturbative and non-adiabatic solution of problems involving a time-periodic Hamiltonian such as ours satisfying $\mathcal{H}(t)=\mathcal{H}(t+T)$ with $T=2\pi/\Omega$.
The Floquet theorem assures that there is a complete set of solutions of the form $\ket{\psi_{\alpha}(t)}=\exp(-i\varepsilon_{\alpha} t) \ket{\phi_{\alpha}(t)}$, where $\varepsilon_{\alpha}$ are the quasienergies and $\ket{\phi_{\alpha}(t+T)}=\ket{\phi_{\alpha}(t)}$ are the Floquet states obeying: 
\begin{equation}
\mathcal{H}_F \ket{\phi_{\alpha}(t)}=\varepsilon_{\alpha} \ket{\phi_{\alpha}(t)},
\end{equation}

\noindent where $\mathcal{H}_F \equiv \mathcal{H}-i \partial/\partial t$ is the Floquet Hamiltonian. Thus, one gets an eigenvalue problem in the direct product space (Floquet or Sambe space~\cite{sambe_steady_1973}) $\mathcal{R} \otimes \mathcal{T}$ where $\mathcal{R}$ is the usual Hilbert space and $\mathcal{T}$ the space of $T-$periodic functions spanned by $\Braket{t|n}=\exp{(i n \Omega t)}$. The index $n$ is often called the \textit{replica} index and can be associated to different photon channels. In this picture, an electron entering lead $\alpha$ in a given reference replica $n_0$ can scatter to lead $\beta$ in replica $m$, thus exchanging $n=m-n_0$ quanta of the time-dependent modulation. Since asymptotically the different replicas are decoupled (as the time-modulation is limited to the sample), the total transmission probability from $\alpha$ to $\beta$, $\mathcal{T}_{\beta, \alpha}(\varepsilon)$, can be obtained by summing the probabilities associated to each of these processes (denoted with $\mathcal{T}^{(n)}_{\beta, \alpha} (\varepsilon)$):

\begin{equation}
\mathcal{T}_{\beta, \alpha}(\varepsilon) = \sum_{n} \mathcal{T}^{(n)}_{\beta, \alpha} (\varepsilon),
\end{equation}

These probabilities can be computed using standard Green's functions techniques~\cite{calvo_non-perturbative_2013,stefanucci_time-dependent_2008,arrachea_relation_2006}. 
In a two-terminal setup in the lineal response regime (small bias voltage), the time-averaged current is given by~\cite{foa_torres_multiterminal_2014}:

\begin{equation}
\overbar{{\cal I}}\simeq \frac{2e^2}{h} {\cal T_{+}}(\varepsilon_F) V + \frac{4e}{h} \int {\cal T_{-}}(\varepsilon)f(\varepsilon)  d\varepsilon.
\label{Floquet-Current-2-terminals-zerobias}
\end{equation}

\noindent where ${\cal T_{\pm}} = ({\cal T}_{R,L} \pm {\cal T}_{L,R})/2$, $f=f_L+f_R$ is the sum of the Fermi distribution functions at each lead ($L$ or $R$) and $V$ is the bias voltage. The second term on the right-hand side corresponds to a \textit{pumping current} arising because of the asymmetry of the transmission coefficients. This current remains even at zero bias voltage (and it may even flow against an applied voltage). Such currents have been studied extensively in the past~\cite{moskalets_floquet_2002,altshuler_pumping_1999,giazotto_josephson_2011,kaestner_non-adiabatic_2015} and more recently have applied to the context of the shift photocurrents problem~\cite{bajpai_spatio-temporal_2019}. As we will see later on, in our case these pumping currents can be tuned by the interplay between the native topological states of the system, the laser polarization and photon assisted processes.

\begin{figure}[hbt!]
\centering
\includegraphics[width=0.85\linewidth]{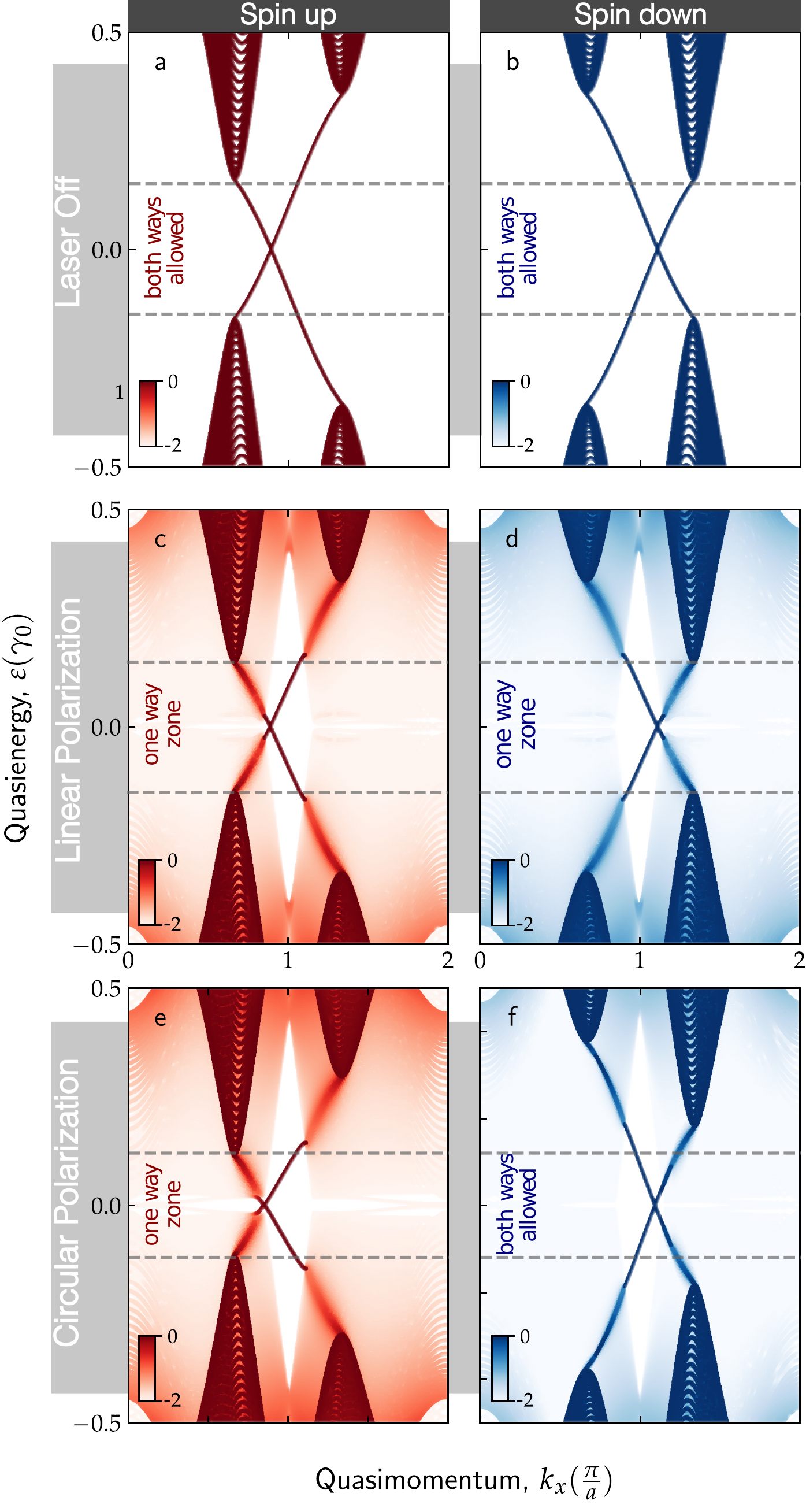}
\caption{Spin resolved bandstructure of a honeycomb ribbon with intrinsic spin-orbit coupling. Computation performed over a zigzag nanoribbon of width $W=100a$ ($\sim 25\, \text{nm}$). Panels (a-b) without irradiation. Panels (c-d) and (e-f) for linear and circular polarized irradiation, respectively. Red(blue) denote spin up(down). We consider $\Delta=0.1$, $\gamma_{\text{SO}}=0.05$, $\hbar \Omega=1.5$ and $z_x=z_y=0.15$. One or both ways allowed transport is highlighted for edge states bridge the energy gap. The color scale in the bottom shows the time-averaged density of states in log-scale.}
\label{fig:bands}
\end{figure}

\textit{Quasi-energy dispersion.} Let us start our discussion by analyzing the dispersion relations for a ribbon of laser-illuminated graphene with spin-orbit coupling and a staggering potential. This is shown in Fig.~\ref{fig:bands} for linear (c-d) and circular (e-f) polarization and also without radiation (a-b). Without radiation, when the spin-orbit term dominates over the staggering one has the expected topological states bridging the gap. The staggering is responsible for the asymmetry between the valleys, while the overall time-reversal symmetry enforces the mirror symmetry between the plots (when exchanging $k$ by $-k$) for the different spin-components. The color scale encodes the contribution of each state to the time-averaged density of states~\cite{oka_photovoltaic_2009} which is given by the weight of each state on the reference replica ($n=0$), which is uniform and equal to unity in absence of radiation. For linear and circular polarization, the lighter tones (notice the log scale) show the regions with states due to the other replicas (each shifted by $\hbar \Omega$). Radiation will introduce a coupling between the replicas (or, in other words, a coupling between a state with a given $k$ at energy $\varepsilon$ and other states at the same $k$ with energy $\varepsilon+n\hbar\Omega$). The effect of such coupling is the hybridization of the native topological states of this system with the continuum provided by the replicas. In the figure this is evidenced as regions where the lines bridging the gap become blurred (the log scale emphasizes these regions which in normal scale will be hardly visible). Later on, we will see how transport is disrupted due to this hybridization.

Notice that the hybridization with the continuum appearing here is different from that studied in Refs.~\cite{foa_torres_crafting_2016,dal_lago_one-way_2017,berdakin_directional_2018} where the continuum is provided by the states of a second layer in bilayer graphene. In contrast, here this is due to coupling with the continuum in other replicas through photon-assisted processes. Furthermore, spin plays a crucial role in the selection rules as we will highlight later on.

When comparing the results for linear and circular polarization in Fig.~\ref{fig:bands} we find an interesting asymmetry: while with linearly polarized light time reversal symmetry is preserved, circular polarization breaks it. The panels for spin up and down in Fig.~\ref{fig:bands}e-f do not mirror each other as when TRS is preserved (panels c-d) and the response is thus expected to be spin-selective. As we will see next, this leads to a deeper selection rule tied to a spin-dependent circular-dichroism effect.

\begin{figure*}[hbt!]
\centering
\includegraphics[width=0.9\linewidth]{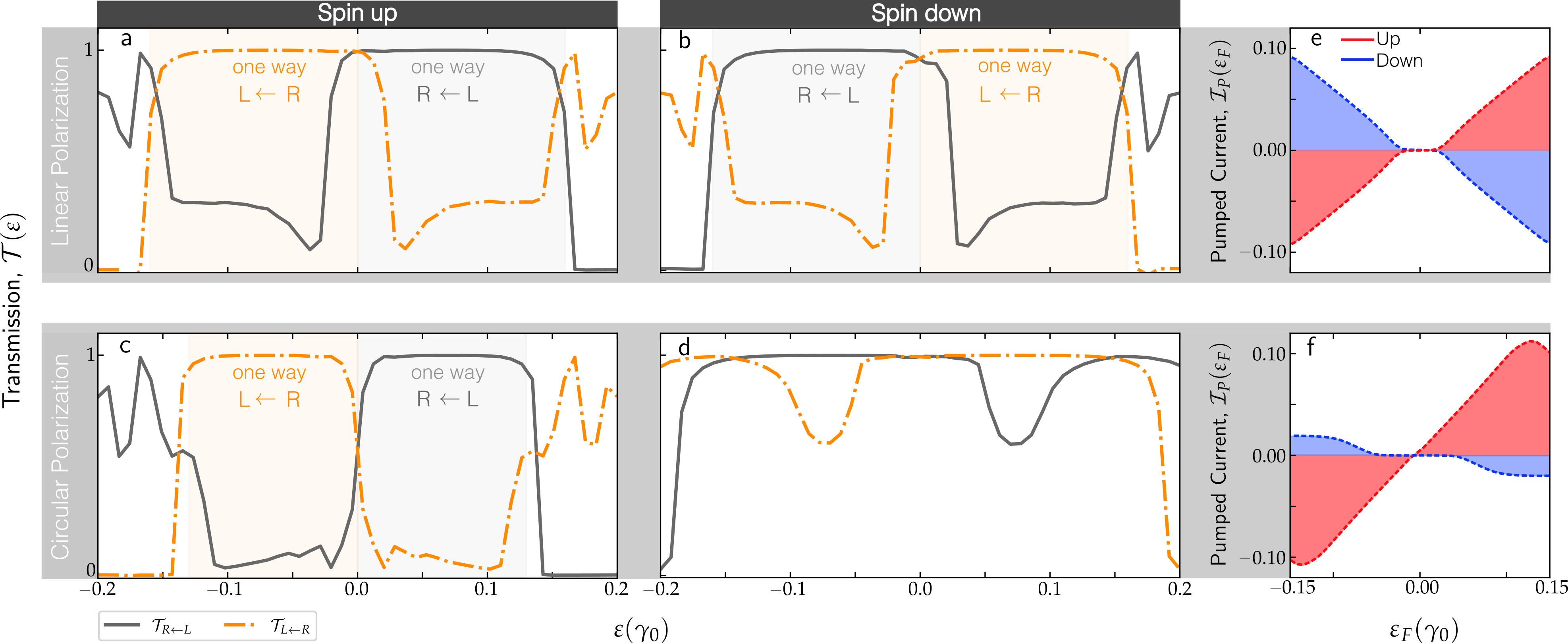}
\caption{Effective transport properties for the scattering process averaged over one irradiation period. Parameters are the same than in Figure ~\ref{fig:bands}. Panels (a-d) show the transmission probabilities within the native gap. One-way charge transport is achieved in (a), (b) and (c) while in (d) no one-way effect is witnessed. Remarkably, the difference between both polarizations can be understood in terms of the circular dichroism on each copy of Haldane model. Panels (e-f) present the spin-resolved pump currents obtained for linear and circular polarization. Due to the presence of pumping currents in Floquet context, this behavior translates into a zero charge pumping with linear polarization yielding pure spin currents (e) while with circular polarizations (f) one can achieve spin polarized charge currents. The former can be tweaked by changing from right-hand to left-hand circular polarization.}
\label{fig:totalT}
\end{figure*}

\textit{Transport properties.--} Let us now turn to the transport properties. We consider a two-terminal setup where a central region is being illuminated while the leads remain in equilibrium. All the Hamiltonian parameters of the scattering zone and leads are equal. By using Floquet scattering theory as mentioned earlier, we compute the total transmission probabilities as a function of the energy of the incident electrons ($\varepsilon$). Furthermore, we can resolve the contributions of both spin components as shown in Fig.~\ref{fig:totalT}a-d (readers can find a detailed comparison between Floquet bandstructure and transport signatures in the supporting information). While without laser illumination one would expect a perfect and reciprocal transmission equal to unity for energies within the bulk gap, here we see a different picture. First, the left-to-right and right-to-left transmission probabilities differ, as is usual in driven systems with broken symmetries. But interestingly, the response is also highly sensitive to the spin component for circularly polarized light: the results show that one spin component gets stronger scattering (deviations from unity) while the other is less affected. This startling difference in the response for different spins (this is, the difference between Figs.~\ref{fig:totalT}c-d) begs for an explanation.

The results for the current, which we can resolve in its spin components are shown in Fig.~\ref{fig:totalT} e (linear polarization and f (circular polarization). A first fact advanced earlier is that because of the non-reciprocity, there is a photo-generated current that appears even at zero bias voltage. This type of pumped current~\cite{moskalets_floquet_2002} or photocurrent in this case~\cite{bajpai_spatio-temporal_2019} is intertwined with the symmetry breaking induced by the different terms in the Hamiltonian. While inversion symmetry is broken in all cases (due to the staggering term), for linear polarization TRS is preserved and hence no net charge current can flow in this spinful case. The current per spin component is non-vanishing as shown in Fig.~\ref{fig:totalT}e, and together both components give a pure spin current.

In contrast, for circular polarization there is a non-vanishing current which turns out to be spin-polarized (see Fig.~\ref{fig:totalT}f). The polarization depends on the Fermi energy, being almost perfect close to the charge neutrality point and of about $83\%$ at higher/lower energies. The spin-selective non-reciprocity (Figs.~\ref{fig:totalT}c-d) under circular polarization together with the spin-polarized photocurrents (Fig.~\ref{fig:totalT}f) are the main numerical results of this paper. Notice that the spin polarization can be inverted by changing the handedness of the laser polarization.

To rationalize the transport results and the quasienergy dispersions we now discuss several points that altogether explain our findings. But first, we need to dig deeper by presenting the different contributions to the total transmissions shown in Fig.~\ref{fig:totalT}. Indeed, an electron entering the illuminated sample with energy $\varepsilon$ can exit elastically (without emitting or absorbing a net number of photons) or inelastically. The partial transmissions ${\cal T}_{i,j}^{(n)}$ for linear and circular polarization are shown in Fig.~\ref{fig:circ} and a discussion of the role of the inelastic back scattering can be found in the supplementary information. The insets of Fig.~\ref{fig:circ} show the transition matrix elements between the unperturbed initial and final states. These insets confirm our previous observation that the propagating state with spin down traversing the device is much less disturbed by circularly polarized light than the other.

\begin{figure*}[hbt!]
\centering
\includegraphics[width=\linewidth]{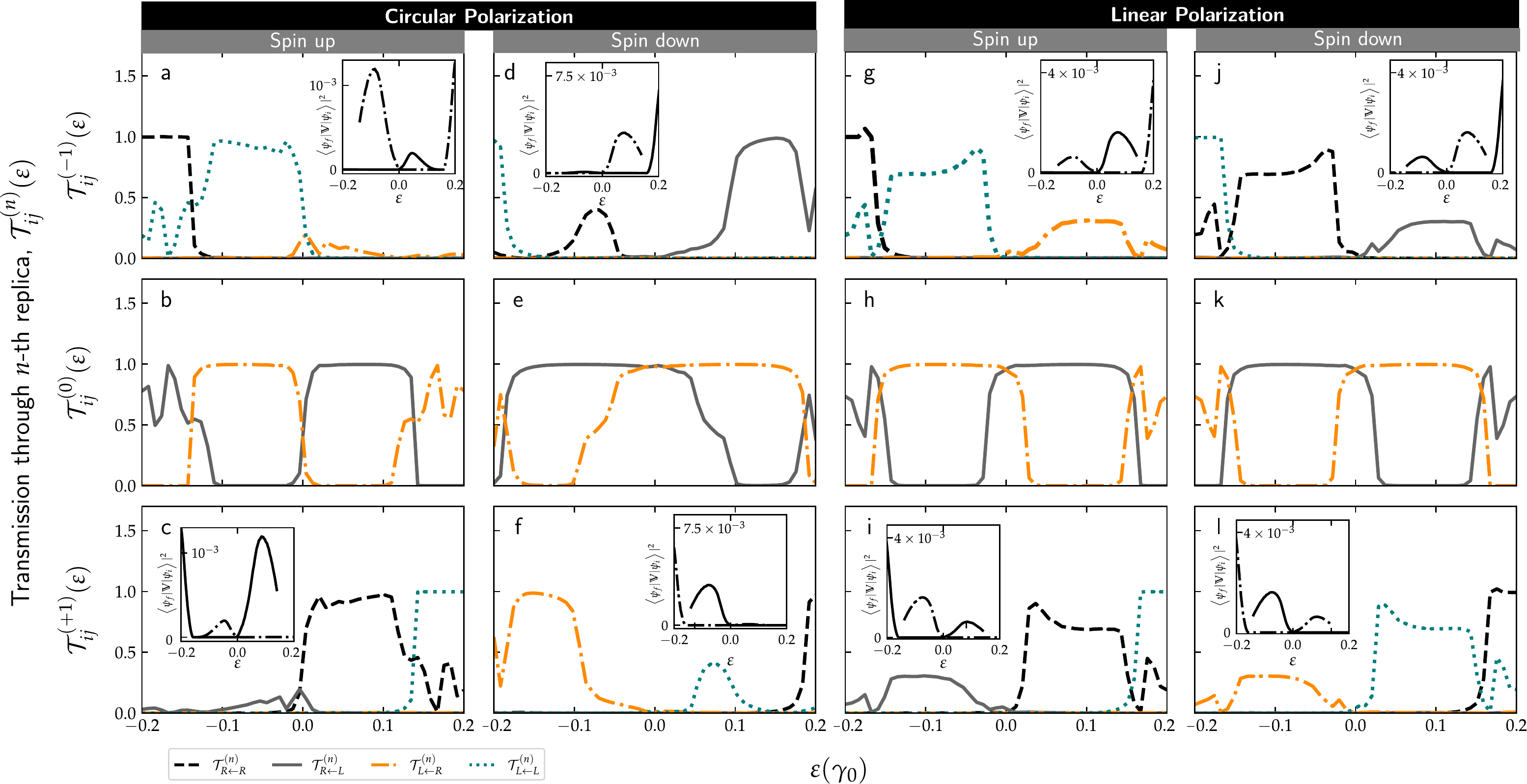}
\caption{Detailed transmission probabilities averaged in time. Here, $\mathcal{T}_{ij}^{(n)}(\varepsilon)$ stands for the transmission through a channel mediated by an exchange of $n$ photons with an electron incident with quasienergy $\epsilon$. From panels (a) to (f) circular polarization transport is shown, describing the whole process for each spin channel independently. From panel (g) to (l) the same information is depicted for linear polarization. Elastic channels prove to be the main source of transmission, while reflection processes, leading to a one-way transport in regions within the bulk gap, are completely mediated by photon-dressed processes, a distinctive signature of hybridization of the states with the continuum. Insets on each panel quantify the degree of coupling of a native topological state and the bulk bands states induced by higher order replicas. In the insets, solid and dashed lines represent opposite edge chiral states.}
\label{fig:circ}
\end{figure*}

The following general points explain the observed numerical features:
\begin{enumerate}
    \item \textit{Generalized symmetry in Floquet space.} The following relation among transmission probabilities is verified in our case: 
\begin{equation}
    \mathcal{T}^{(n)}_{\beta, \alpha} (\varepsilon)=\mathcal{T}^{(-n)}_{\alpha,\beta} (-\varepsilon).
\end{equation}
This is enforced by an underlying symmetry in Floquet-space:  
\begin{equation}
    \Gamma \mathcal{H(\textbf{k})} \Gamma^{\dagger} = -\mathcal{H(\textbf{k})}^{*},
    \label{sym}
    \end{equation}
    where $\Gamma = \sigma_y K$, $K$ being the complex conjugation operator. This generalized symmetry thus inverts the energy sign while mirroring the space and replica coordinates. This symmetry, which is fulfilled in our device setup, therefore links the transmission probabilities in the different panels of Figs.~\ref{fig:totalT} and~\ref{fig:circ}, which also serves as a numerical test of our results. We note that the similar relations have also been used in a different context in Floquet systems~\cite{balabanov_transport_2019}. 
    \item \textit{Spin-selective dichroism effect: a selection rule linking Chern number, circular polarization handedness and spin.} Under illumination with circularly polarized light we observe a marked transport asymmetry between spin components, and also within the same spin subspace when the handedness of the laser polarization changes. The latter is commonly referred to as circular dichroism. The existence of circular dichroism in the presence of both a complex next-nearest-neighbor coupling and a staggering potential for the bulk states has been discussed in Ref.~\cite{ghalamkari_perfect_2018}. In that reference, Ghalamkari and coauthors find that there is a selection rule which ties the Chern number of the topological phases found in the Haldane model~\cite{haldane_model_1988} with a distinctive response to left and right circular polarization. 
    
    In our case, when looking at each spin component separately, our numerical results show that this selection rule persists for a finite system, when the transitions include an edge state and a state in the bulk spectrum. Furthermore, the fact that both spin components are related by time-reversal in ${\cal H}_0$ produces an inversion of the circular dichroism when passing from spin up to spin down, since it follows the sign inversion of the spin-resolved Chern number. Here we observe that this inversion of the circular dichroism is also fulfilled for the finite system, a fact which one might intuitively tie in with the bulk-boundary correspondence.
    
    \item \textit{Hybridization of edge states with the continuum provided by a different Floquet replica.} The selection rule stated in point 2 plays a crucial role in establishing the possible light-induced transitions among the electronic states. For Fermi energies within the bulk gap of the sample, thanks to photon-assisted processes, the topological edge states at $\varepsilon$ can now transition towards the continuum of states at $\varepsilon+n\hbar\Omega$ (where the system is ungapped). Based on point 2, this hybridization with the states of a different replica is expected to be insensitive to the spin for linear polarization but not for circular polarization. This is verified by numerically computing the modulus squared of the matrix element of the perturbation among the initial and final states, see insets of Fig.~\ref{fig:circ}. 
    
\end{enumerate}

 For our numerical simulations we employed a general model with staggering potential and spin-orbit coupling  compatible with Germanene and Stanene, but we notice that the interplay between the staggering strength, the spin-orbit coupling and the laser frequency allows for a broad range of materials where the predicted photocurrents could be observed. Indeed, we require a system with  with broken inversion symmetry hosting topological states. Within the Kane-Mele model this means that $2\Delta/\gamma_{SO}<3\sqrt{3}$~\cite{kane_quantum_2005,haldane_model_1988}. Furthermore, for the hybridization of the topological states with the continuum of the Floquet replicas to occur, the photon energy needs to be not smaller than the bulk gap of the non-irradiated system and not so large so that there are no continuum states $\hbar\Omega$ above a given energy. Fortunately, these conditions do not impose a restriction within the experimentally relevant regime of laser frequencies spanning from the mid-infrared to the visible range (see supplementary information). On the other hand, the temperature needed for the experimental realization and  the fine-tuning of the irradiation condition will  depend on the energy gap of the unperturbed material. The required laser intensities are smaller than those required to observe Floquet-Bloch states, as here we need sufficient coupling with a continuum of states. From our numerics, for typical mid-infrared wavelengths ($\sim 160$ meV) we estimate that intensities in the range of $1-10$ mW/$\mu\mathrm{m}^2$ would suffice.

Let us now discuss the influence of Rashba spin-orbit coupling. A Rashba term introduces spin-flip processes and it is a legitimate source of concern. Our numerical results evidence robustness of the photocurrents against this term (see supplementary information). This is because the mechanism relies on the existence of topologically protected states in the non-irradiated material, which are originated by intrinsic SOC and which extend to a region of parameters with moderate values of Rashba SOC. Indeed, the topological states are robust against a moderately strong Rashba SOC ~\cite{kane_quantum_2005}: for $\gamma_{R}<2\sqrt{e}\gamma_{SO}$ ($\gamma_{R}$ and $\gamma_{SO}$ being the strengths of the Rashba and intrinsic spin-orbit coupling terms) the resulting phase diagram is adiabatically connected to the quantum spin-hall phase of the Kane-Mele model. This topological protection is expressed in the fact that the matrix element between two topological counter-propagating states at one edge of any perturbation that preserves TRS is zero. In our case, circular polarization does not preserve TRS but rather than introducing a matrix element between counter-propagating edge states, leads to a selective hybridization of the edge states with the continuum at an energy differing by the photon energy from them. This is why our results are robust against spin non-conserving terms over a broad range of parameters.

\textit{Final remarks.--} 
Using Floquet scattering theory we show how laser illumination can selectively disrupt the edge states of a two-dimensional topological insulator depending on their spin. This selectivity, which stems from the interplay between a spin-selective selection rule together with the hybridization of the edge states with the continuum of another Floquet replica, manifests by the generation of pure spin currents and spin-polarized charge photocurrents under linearly and circularly polarized laser-illumination, respectively. We emphasize that, in both cases, the spin and charge currents flow even at zero-bias voltage. Furthermore, the direction and spin polarization of these currents can be tuned by changing the incident electronic energy and the handedness of light polarization, thereby providing an experimental handle to control photocurrents. In this sense, given the generality of our model, we expect for the photocurrents predicted here to be experimentally accessible in two-dimensional materials by using laser illumination in the mid-infrared.

\begin{acknowledgments}
E.A.R.-M. acknowledges support by the ANID (Chile)PFCHA,  DOCTORADO  NACIONAL/2017,  under  ContractNo.  21171229. We thank the support of FondeCyT (Chile) under grant number 1170917, and by the EU Horizon 2020 research and innovation program under the Marie-Sklodowska-Curie Grant Agreement No. 873028 (HYDROTRONICS Project). L. E. F. F. T. also acknowledges the support of The Abdus Salam International Centre for Theoretical Physics and the Simons Foundation. 
\end{acknowledgments}


%
\end{document}


\newpage

\section{A detailed comparison between Floquet bandstructure and transport signatures}
\begin{figure}[h]
  \centering
  \includegraphics[width=1.0\linewidth]{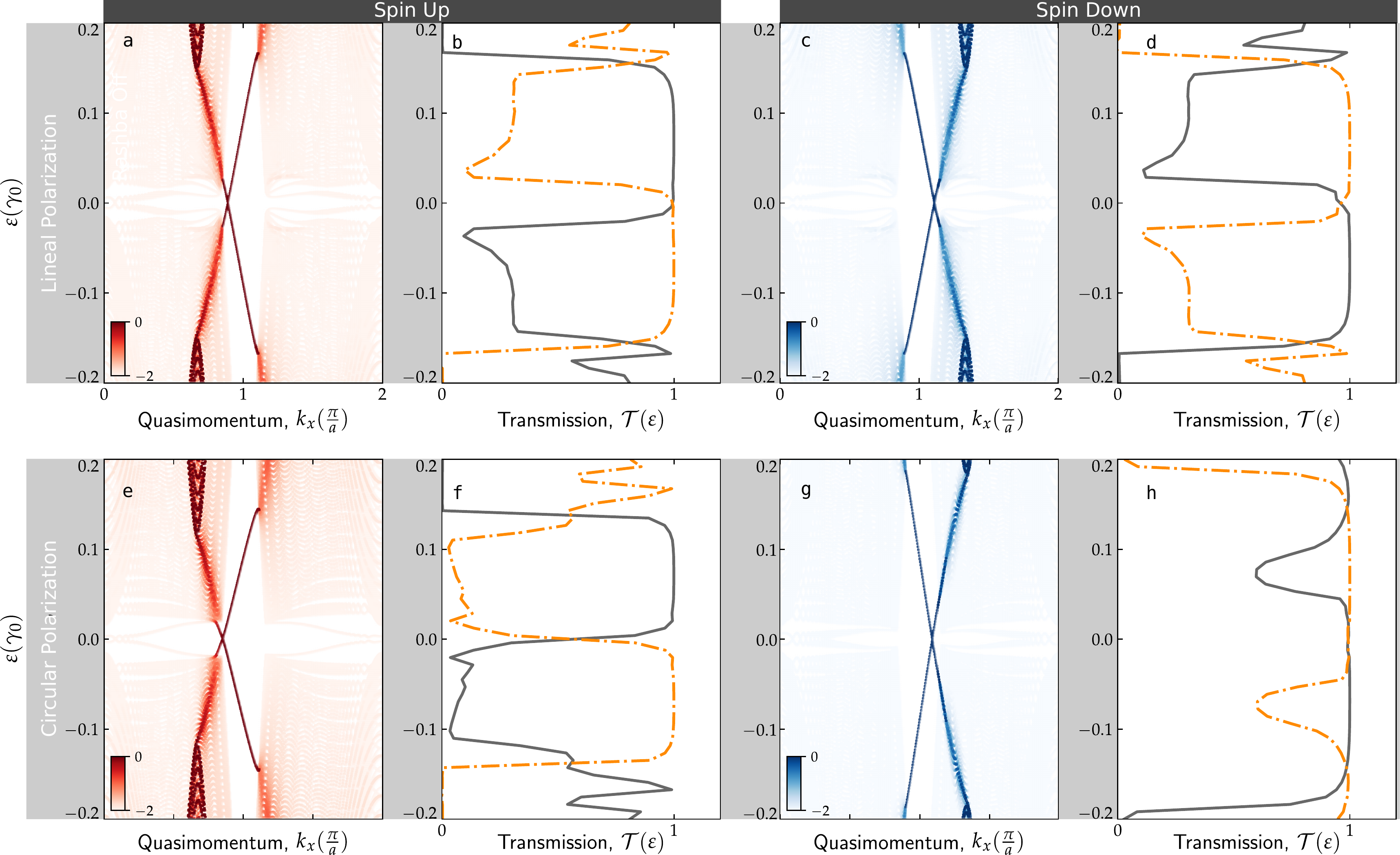}
    \caption{Each two-fold panel shows the quasi-energies versus quasi-momemtum (left) with a color scale encoding the weight on the zero replica, and the transmission probabilities from right to left (dash-dotted orange line) and from left to right (black solid line) on the right.}
  \label{fgr:BandsVsTij}
\end{figure}

The correspondence between the band structure and the transport results (including the asymmetry between the transmission probabilities from left to right and from right to left) can be better appreciated in figure ~\ref{fgr:BandsVsTij} which uses the data in Figs. 2e-f  and 3c-d  of the main article. 

One can see that within the bulk gap the topological states for spin-up and down show almost perfect transmission except for the regions where the laser plays a role by hybridizing with a continuum (seen as blurred lines in the quasi-energy spectra).

However, as soon as one the energies are close to the edge of the bulk gap or those regions spanning the bulk bands, more fluctuations arise as we get into the non-topological part of the spectrum. These fluctuations originate from scattering at the interface between the irradiated and non-irradiated areas and different matching problems between sample and leads (even when both are the same material, one is irradiated and the other not). This type of behavior is commonplace in transport simulations and experiments, note for example the celebrated transport experiments by Molenkamp \cite{konig_quantum_2007} on 2D Topological insulators.

Anticipating the problem of matching with leads made of different materials, as present in the alluded experiments,\cite{konig_quantum_2007} and taking advantage that the photocurrents are generated by the laser spot which might be smaller than the sample itself, one can envision an experiment where the photocurrents flow within the same material before reaching the leads far away. In this situation the matching with leads made of different materials is lessened and one recovers a cleaner response. 

Another strategy that could be pursued experimentally to separate contributions to the photocurrents coming from different sources, e.g. the ones predicted here and others of thermal origin, is to exploit that our predicted photocurrents depend strongly on the laser polarization while thermal contributions do not. This has been successfully used for example in  reference \citenum{mciver_light-induced_nature} to highlight the laser-induced Hall effect.

\section{Effect of inelastic backscattering}

\begin{figure}[h]
  \centering
  \includegraphics[width=1.0\linewidth]{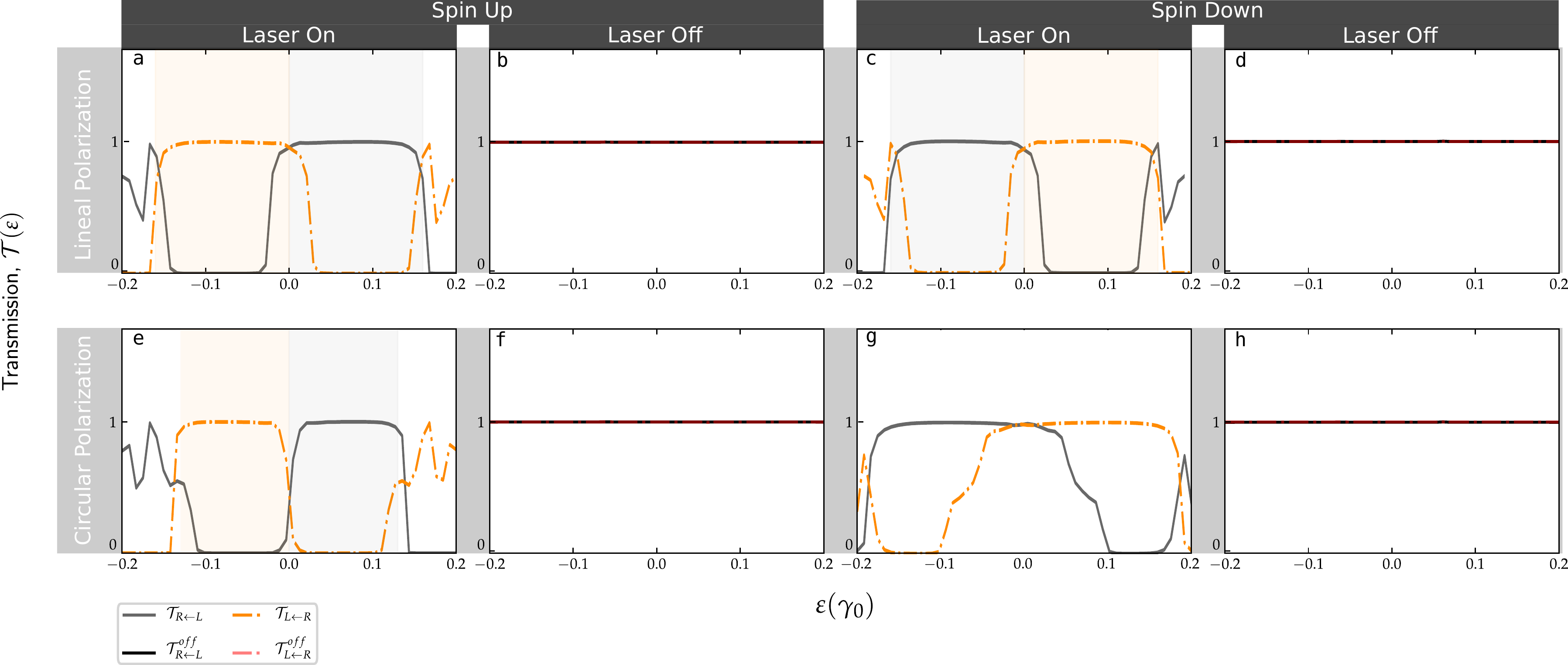}
    \caption{Each two-fold panel shows the transmission probabilities for each spin component with and without the laser. Only elastic components of the transmission are shown. The parameters are the same as in Fig. 4 of the main manuscript}
  \label{fgr:backscatt}
\end{figure}

The hybridization with the continuum provided by the coupling with the laser allows for inelastic processes (processes involving the absorption or emission of energy quanta) which rather than contributing to the transmission tend to suppress it in favor of  backscattering (reflection). Although at first sight it might seem that the inelastic processes influence very little the transmission (from L to R and vice versa), it is actually the opposite: over a large energy span inelastic processes almost fully suppress the transmission. For example, for negative energies Fig. 4a of the main text shows almost unity backscattering on the left lead (dotted blue line) with a concomitant nearly zero elastic transmission from left to right in the same energy range, as shown in Fig. 4b (dark full line). This information is replotted in figure \ref{fgr:backscatt} of this document which also shows the results with the laser off for comparison.

We emphasize that the magnitude of the effect is quite unprecedented, since even moderate laser intensities are able to induce backscattering with almost unit probability over a wide energy range.

\section{The interplay between laser frequency and other relevant parameters}

The effect reported in our manuscript relies on two crucial points: 
\begin{itemize}
    \item First we need an undriven system with broken inversion symmetry hosting topological states. Within the Kane-Mele model this immediately puts a constraint on the relative magnitude of the staggering and the spin-orbit coupling, since a large staggering will lead to a transition to a topologically trivial phase after a gap closing. The topological phase appears for $2\Delta/\gamma_{SO}<3\sqrt{3}$~\cite{kane_quantum_2005,haldane_model_1988}. 
    \item Furthermore, for the effect to be observable one would need a sizable gap of several $k_{B}T$. Then we need to introduce the hybridization of the topological states with the continuum provided by the Floquet replicas. For this to occur we need for the frequency to be not smaller than the bulk gap of the non-irradiated system and not so large so that there are no continuum states $\hbar\Omega$ above a given energy. Fortunately, these conditions do not impose a restriction within the experimentally relevant regime of laser frequencies spanning from the mid-infrared to the visible range. 
\end{itemize}

\begin{figure}[h]
  \centering
  \includegraphics[width=0.8\linewidth]{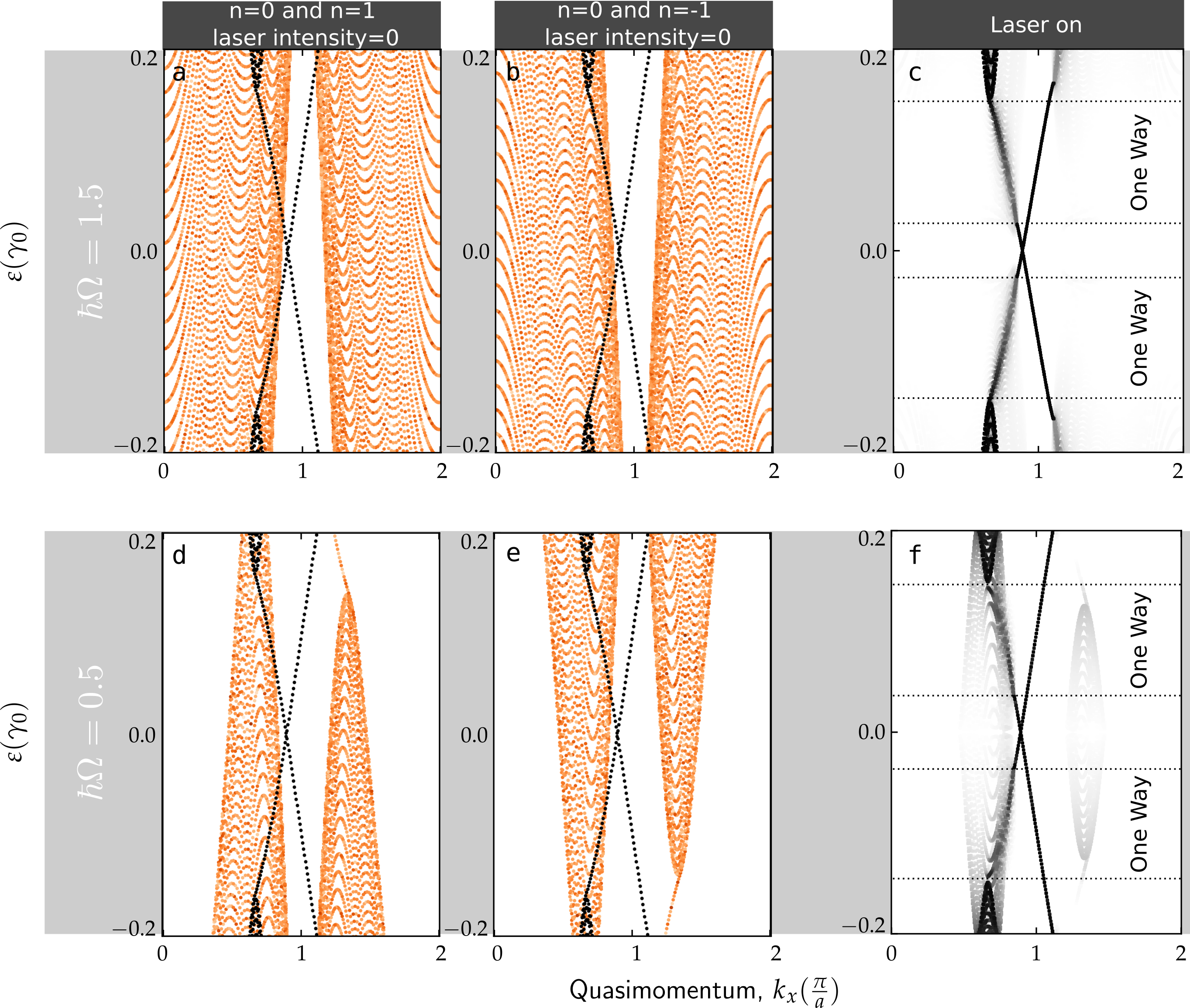}
    \caption{Non-irradiated $n=0$ (black) and $n \pm 1$ replicas (orange) bandstructures  together with the corresponding Floquet bandstructure obtained when the laser is on. Panels (a-c)  for $\hbar\Omega = 1.5$ and (d-f) for $\hbar\Omega = 0.5$}
  \label{fgr:BandsVsfreq}
\end{figure}

The latter can be appreciated by plotting the non-irradiated bandstructure together with those of the replicas (just the shifted bandstructure) and checking that geometrically the replicas provide a background of states with which the edge states can hybridize. This is shown in figure \ref{fgr:BandsVsfreq} of this document (left and center panels). To further support our argument the full Floquet bandstructure is shown in the right panel. 

\section{Effect of Rashba spin-orbit coupling}

The scattering region, described in the main text, can be generalized by adding a Rashba spin-orbit term.\cite{KM-Z2} This term typically arises because of a breaking the mirror symmetry and couples nearest neighbors. Therefore, the Hamiltonian now reads 
\begin{align}
\label{Hamiltonian}
\mathcal{H}_{0} & =\sum_{i,s_{z}}E_{i}c_{i.s_{z}}^{\dagger}c_{i,s_{z}}-\gamma_{0}\sum_{\left\langle i,j\right\rangle }c_{i,s_{z}}^{\dagger}c_{j,s_{z}}\\
 & -i\gamma_{\text{SO}}\sum_{\left\langle \left\langle i,j\right\rangle \right\rangle }\nu_{i,j}s_{z}c_{i,s_{z}}^{\dagger}c_{j,s_{z}}+i\gamma_{\text{r}}\sum_{\left\langle i,j\right\rangle }c_{i,s_{z}}^{\dagger}\left(\mathbf{s}\times\hat{\mathbf{d}}_{ij}\right)c_{i,s_{z}}
\end{align}

\begin{figure}[ht]
  \centering
  \includegraphics[width=1.0\linewidth]{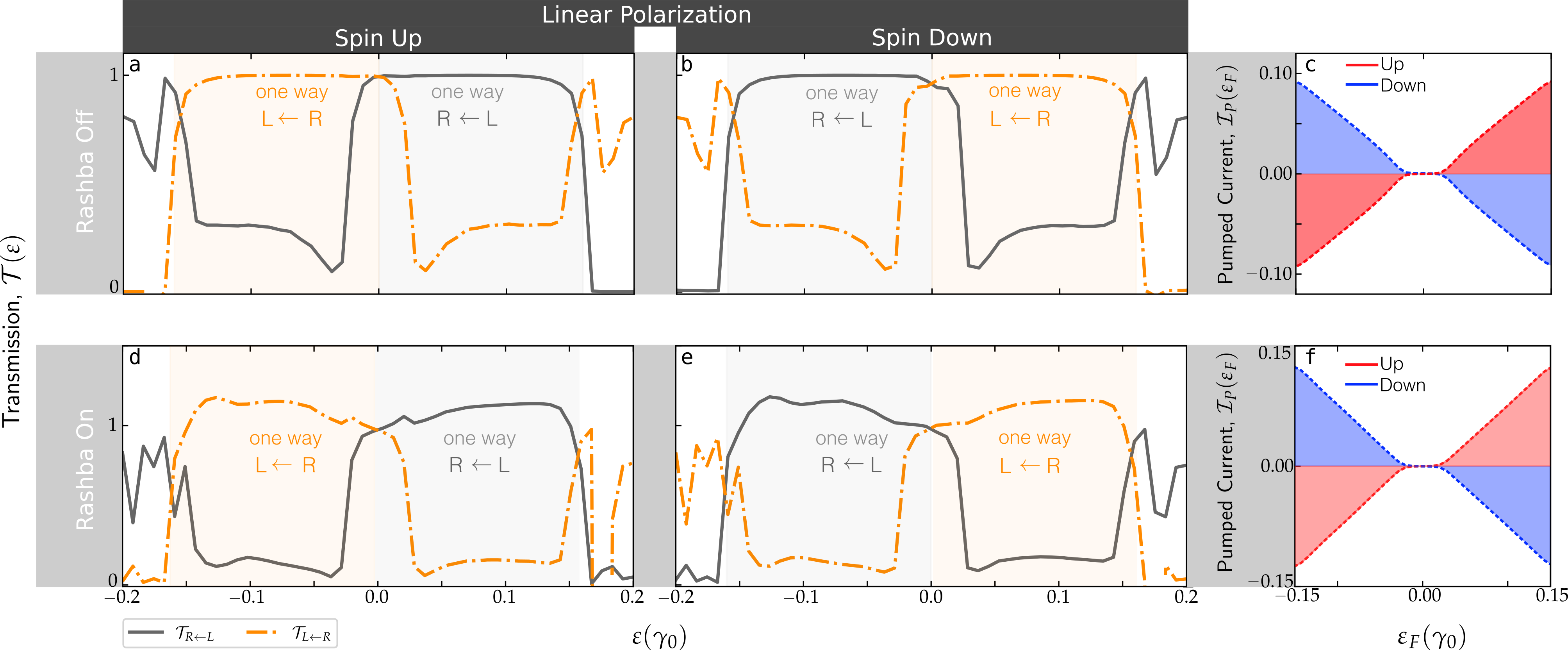}
  \caption{Two terminal total transmission probability and pumped currents for linear polarization and both spin projections in direction $\hat{z}$. The panels (a-c) show the case when the Rashba term $\gamma_\text{r}$ is set to 0. Panels (d-f) show the effect of mirror symmetry breaking due to Rashba spin-orbit coupling $\gamma_{\text{r}}=0.0075$, alike as heavy materials such as germanene. As in the main text we have set the energy scale such that $\gamma_0=1$ and we consider $\Delta=0.1$, $\gamma_{\text{SO}}=0.05$, $\hbar \Omega=1.5$ and $z_x=z_y=0.15$.}

  \label{fgr:rashba_lin}
\end{figure}

\begin{figure}[ht]
  \centering
  \includegraphics[width=1.0\linewidth]{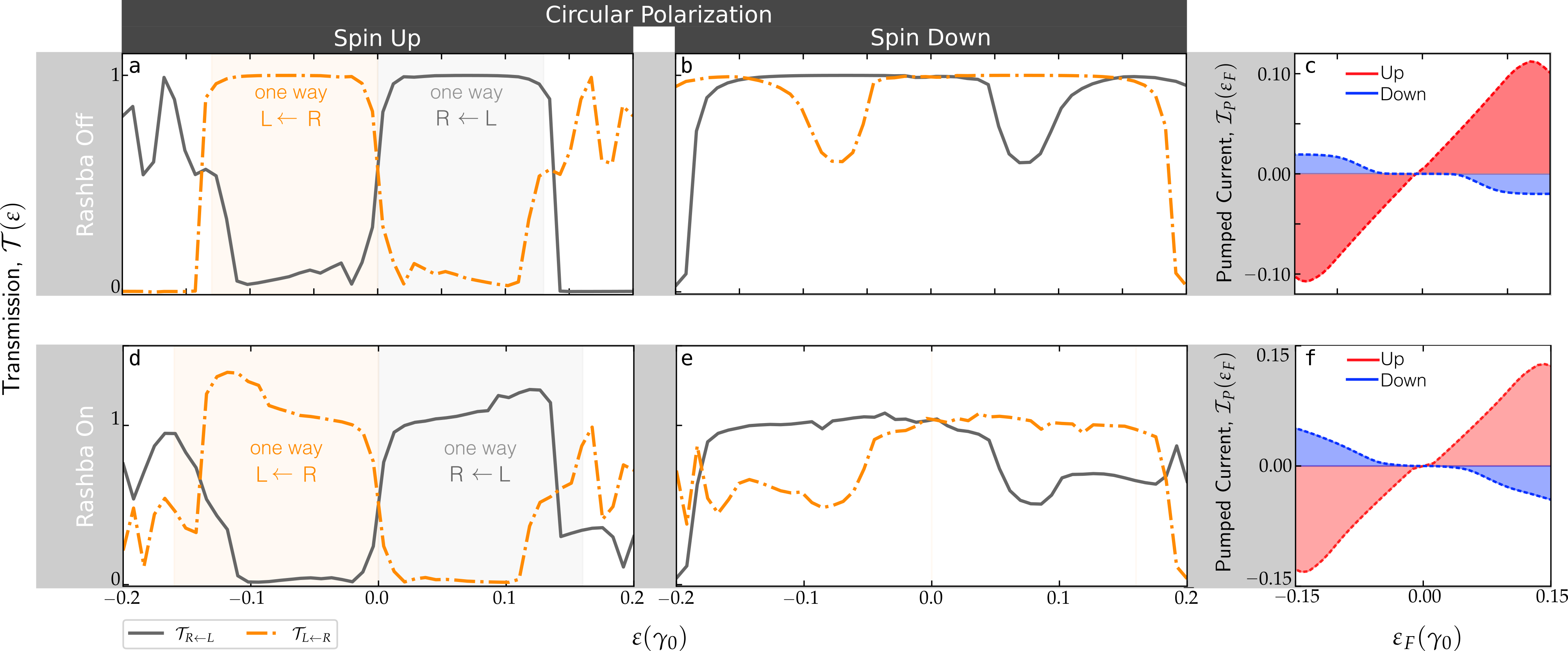}
     \caption{Two terminal total transmission probability and pumped currents for circular polarization and both spin projections in direction $\hat{z}$. The panels (a-c) show the case when the Rashba term $\gamma_\text{r}=0$. Panels (d-f) show the effect of mirror symmetry breaking due to Rashba spin-orbit coupling $\gamma_{\text{r}}=0.0075$, alike as heavy materials such as germanene. As in the main text we have set the energy scale such that $\gamma_0=1$ and we consider $\Delta=0.1$, $\gamma_{\text{SO}}=0.05$, $\hbar \Omega=1.5$ and $z_x=z_y=0.15$.}
  \label{fgr:rashba_circ}
\end{figure}

\noindent with $c_{i, s_z}^{\dagger}$ and $c_{i, s_z}^{{}}$ are the creation and annihilation operators for electrons at the $\pi$-orbital on site
$i$ with spin up $s_z=1$ or spin down $s_z=-1$. $\gamma_{0}$ is the nearest-neighbors hopping amplitude and $\gamma_{\text{SO}}$ is the intrinsic spin-orbit coupling. $\nu_{i,j}$ is $+1$ ($-1$) if the path from $j$ to $i$ is clockwise (anticlockwise). The on-site energies $E_{i}$ are chosen equal to $\Delta$ ($-\Delta$) for the sites on the $A$ ($B$) sublattice. The last term introduces Rashba spin-orbit coupling with amplitude $\gamma_{\text{r}}$, where $\textbf{s}$ is a vector with the Pauli matrices as components and $\hat{\mathbf{d}}_{ij}$  denotes first neighbors hopping vector connecting the sites $i$ and $j$,  hence, this term explicitly mixes the spin degrees of freedom, making $s_z$ no longer a good quantum number.

Figures S1 and S2 of this supplementary material show our numerical results including the Rashba term demonstrating the robustness of the predicted photocurrents. Note that results with Rashba spin-orbit coupling show transmission probabilities per spin component which might be greater than 1, this does not violate unitarity as the total transmission (the sum over the two spin components) remains bounded by the number of channels in this energy range (2) and shows that the Rashba term induces a conversion between spin up and down.

\bibliography{2dTI_floquet.bib}